\documentclass[useAMS,usenatbib,usegraphicx]{mn2e}

\newcommand{\avg}[1]{\langle #1 \rangle}
\def\la{\mathrel{\hbox{\rlap{\hbox{\lower4pt\hbox{$\sim$}}}\hbox{$<$}}}}

\usepackage{epsf}
\usepackage{amssymb,amsmath}

\title[Limits on the power law profiles of elliptical galaxies]
{Limits on the power-law mass and luminosity density profiles of elliptical galaxies from gravitational lensing systems
}

\author[Cao et al.]
{Shuo Cao$^{1}$, Marek Biesiada$^{1,2}$, Meng Yao$^{1}$, and Zong-Hong Zhu$^{1}$\thanks{zhuzh@bnu.edu.cn}\\
$^1$ Department of Astronomy, Beijing Normal University, 100875, Beijing, China; \\
$^2$ Department of Astrophysics and Cosmology, Institute of Physics, University of Silesia, Uniwersytecka 4, 40-007 Katowice, Poland
}

\begin{document}

\date{\today}

\voffset- .5in

\pagerange{\pageref{firstpage}--\pageref{lastpage}} \pubyear{}

\maketitle

\label{firstpage}

\begin{abstract}
We use 118 strong gravitational lenses observed by the SLACS, BELLS, LSD
and SL2S surveys to constrain the total mass profile and the profile
of luminosity density of stars (light-tracers) in elliptical
galaxies up to redshift $z \sim 1$. Assuming power-law density
profiles for the total mass density, $\rho=
\rho_0(r/r_0)^{-\alpha}$, and luminosity density, $\nu=
\nu_0(r/r_0)^{-\delta}$, we investigate the power law index and its
first derivative with respect to the redshift. Using Monte Carlo
simulations of the posterior likelihood taking the Planck's
best-fitted cosmology as a prior, we find $\gamma= 2.132\pm
0.055$ with a mild trend $\partial \gamma/\partial z_l= -0.067\pm
0.119$ when $\alpha=\delta=\gamma$, suggesting that the total
density profile of massive galaxies could have become slightly
steeper over cosmic time. Furthermore, similar analyses performed on
sub-samples defined by different lens redshifts and velocity
dispersions, indicate the need of treating low,
intermediate and high-mass galaxies separately.
Allowing $\delta$ to be a free parameter, we obtain $\alpha=
2.070\pm0.031$, $\partial \alpha/\partial z_l= -0.121\pm0.078$, and
$\delta= 2.710\pm0.143$. The model in which mass traces light is
rejected at $>95\%$ confidence and our analysis robustly indicates
the presence of dark matter in the form of a mass component that is
differently spatially extended than the light. In this case,
intermediate-mass elliptical galaxies ($200$ km/s $ <
\sigma_{ap} \leq 300$ km/s) show the best consistency with the
singular isothermal sphere as an effective model of galactic lenses.
\end{abstract}

\begin{keywords}
gravitational lensing: strong - galaxies: structure - cosmology: theory
\end{keywords}

\section{Introduction}\label{sec:introduction}

Since the discovery of the first strong gravitational lens system
Q0957+561 \citep{1979Natur}, strong lensing has developed into an
important astrophysical tool suitable for investigating both background
cosmology \citep{Zhu00a,Zhu00b,Cha03, Cha04, Mit05, Zhu08a, Zhu08b}
and the structure and evolution of galaxies \citep{Zhu97,MS98,Jin00,Kee01,KW01,Ofek03,Treu06a}.

A well-defined sample of strong lensing systems with accurately
measured image separations and known redshifts of the lens
and of the source could be useful to test cosmological parameters
such as the present-day matter density $\Omega_m$, cosmic
equation of state etc. \citep{Cha03,Mit05} as well as the
statistical properties of lensing galaxies e.g. stellar velocity
dispersion function or galaxy evolution \citep{CM03,Ofek03}.
Concerning cosmological applications, the first method used for this
purpose was of statistical nature. Essentially, the idea was to
apply the velocity dispersion function (VDF) of early-type galaxies
derived from the SDSS Data Release 5 \citep{Sheth03} in
order to analyze the distribution of observed image separations in
the sample of gravitationally lensed systems taken from Cosmic Lens
All-Sky Survey (CLASS) \citep{Mit05,Zhu08a,Cao11a}, in combination
with semi-analytical modeling of galaxy formation and evolution
\citep{CM03}. Next, the idea of using Einstein radius measurements
of strong lensing systems combined with spectroscopic data (stellar
velocity dispersions) provides another interesting possibility to
set limits on cosmological parameters including the cosmic equation
of state \citep{Biesiada06,Grillo08,Biesiada10,Biesiada11,Cao15}.

In the usual formulation of strong gravitational
lensing, one often approximates the actual lens by an idealized model with
a definite radial mass density profile. Most of lensing galaxies
are massive early-type ellipticals as they dominate the lensing
cross-sections due to their large masses.
Spherically symmetric mass distribution leading to a circularly
symmetric surface mass density and having analytical properties, has
always been assumed in statistical studies of strong lensing systems.
Such classical lens models include singular isothermal sphere (SIS)
and Navarro, Frenk and White (NFW) profile
\citep{Bartelmann96,Navarro96}. In particular, the SIS model with
homologous profile ($\rho_{tot}\sim 1/r^2$) for the total
(luminous+dark) density distribution, was extensively used in the
previous lensing-based cosmological studies
\citep{Fukugita92,Kochanek96,Helbig99}. This simple but well-tested mass model
is strongly supported by recent observations of early-type galaxies
both in the inner regions i.e. within the Einstein radii
\citep{Lagattuta10,Ruff11} and at larger distances of $\sim 50-300
~h^{-1}$ kpc \citep{she04,man06}.

By assuming the isothermal mass-density profile in elliptical
galaxies acting as lenses, the study of \citet{Grillo08} reported
the value for the present-day matter density lying between 0.2 and
0.3 at 99\% confidence level. This early result is consistent with
most of the current data: precision measurements of Type Ia
supernovae \citep{Amanullah10}, BAO cosmological distance ratios
from SDSS galaxy sample \citep{Padmanabhan12}, and the anisotropies
in the cosmic microwave background radiation
\citep{Hinshaw12,Planck1}. Currently, there is no strong and
convincing evidence for deviations from the concordance $\Lambda$CDM
model which invokes cosmological constant playing the role of an
exotic component called dark energy responsible for more than 70\%
of the total energy of the universe. Latest exploration of the
properties of dark energy with different astrophysical probes
carried out by
\citet{Cao11c,Cao12a,Cao12b,Cao12c,Cao12d,Cao13,Cao14} gave the
results in agreement with $\Lambda$CDM. In light of phenomenological
success of the concordance cosmological model, we may use
gravitational lenses for a different purpose: to study the structure
of galaxies (their total mass and luminosity profiles) safely
assuming that the cosmological model based on current
precise observations is reliable.

The deflection of light due to gravitational lensing is sensitive to
the total mass of structures in the universe, independent of their
nature or dynamical state. Therefore, strong gravitational lensing
provides a valuable tool for measuring the mass distribution of
early-type galaxies. As it is known from the gravitational lensing
theory, once the critical density of the system is determined by
cosmology, mass enclosed inside the Einstein radius can be directly
obtained from the measurement of Einstein radius. On the other hand,
by spectroscopically measuring the central stellar velocity
dispersion and assuming some functional form of the mass density
profile of elliptical galaxies, this mass can also be obtained
through a model-dependent dynamical analysis.

Considering a more general power law model for the total mass
density profile of lensing galaxies ($\rho_{tot}\sim 1/r^\gamma$),
\citet{Grillo08} presented a joint gravitational lensing and
stellar-dynamical analysis of 11 early-type galaxies taken from the
Sloan Lens ACS Survey (SLACS), and found that the total density
profile of the lenses (ellipticals) is indeed well approximated by
an isothermal distribution ($\gamma \sim 2$), independent of the
cosmological model adopted. Using newly measured redshifts and
stellar velocity dispersions from Keck spectroscopy, \citet{Ruff11}
analyzed 11 early-type galaxies (median lens redshift $z_l=0.5$)
from Strong Lenses in the Legacy Survey (SL2S). For a fixed
cosmological model ($\Lambda$CDM), they derived the total mass
density slope inside the Einstein radius for each lens, with an
average total density slope $\gamma=2.16\pm0.09$ for their sample.
Using a combined set of SL2S, SLACS, and the Lenses Structure and
Dynamics (LSD) survey data, a mild trend of the cosmic evolution of
$\alpha$ was also detected, with magnitude $\partial \gamma/
\partial z_l=-0.25^{+0.10}_{-0.12}$.
Similar analyses aimed at establishing the evolution of mass-density
profile have also been attempted in the past with negative results.
For example, \citet{Koopmans06a} using the joint SLACS/LSD sample of
massive lens galaxies (with velocity dispersions $\sigma_{ap}>200$
km/s),  concluded that there was no significant evolution of the
total density slope inside one effective radius ($\partial \gamma/
\partial z_l=0.23\pm0.16$). More recently, however
\citet{Bolton12} presented an analysis of a combined sample of SLACS
and BELLS lenses with the conclusion that galaxies at lower
redshifts tend to have steeper mass profiles at a later cosmic times
($\partial \gamma/
\partial z_l=-0.60\pm0.15$). This is consistent with the newest results obtained
by \citet{Sonnenfeld13} from enlarged sample observed by SL2S
combined with SLACS ($\partial \gamma/
\partial z_l=-0.31\pm0.10$).

Given the previous literature listed above
\citep{Koopmans06a,Grillo08,Ruff11,Bolton12,Sonnenfeld13}, the
primary motivation of this work is to develop a reliable
phenomenological model of early-type lensing galaxies suitable for
further cosmological studies. It is only quite recently when
reasonable catalogs of strong lenses: containing more than 100
lenses, with spectroscopic as well as astrometric data, obtained
with well defined selection criteria are becoming available. The
purpose of this work is to use the new sample of 118 lenses
\citep{Cao15} compiled from SLACS, BELLS, LSD, and SL2S to provide
independent constraints on the mass distribution of early-type
galaxies. We consider a general power-law mass density profile
allowing the power-law index to evolve with redshift. Moreover, we
also relax the rigid assumption that the stellar luminosity and
total mass distributions follow the same power-law. In Section 2 we
briefly describe the methodology of subsequent analysis, the sample
and its construction. The results obtained with the full sample and
on several sub-samples are presented in Section 3. The main
conclusions are summarized in Section 4. Throughout this paper we
assume flat $\Lambda$CDM cosmology based on the recent Planck
observations, with $\Omega_m=0.315$ and $h=0.673$ \citep{Planck1}.

\section{Method and data}\label{sec:data}

From the theory of gravitational lensing it follows that, for a
specific strong lensing system, multiple images can only form close
to the so-called Einstein ring $\theta_E$ \citep{Schneider92}.
In the case of circularly symmetric mass distribution, the
mass $M_E$ enclosed within a cylinder of a radius equal to the
Einstein radius is also directly related to the geometry of the
Universe (through angular-diameter distances):
\begin{equation}
\theta_E = \left(\frac{4G M_E}{c^2} \frac{D_{ls}}{D_s D_l}
\right)^{1/2} ~~~.
\end{equation}
Here, $D_s$ is the angular-diameter distance to the source, $D_l$ is
the distance to the lens, and $D_{ls}$ is the distance between the
lens and the source. Throughout this paper we will use angular-diameter distances, unless stated otherwise.

Rearranging terms using $R_E = D_l \theta_E$
($R$ is the cylindrical radius perpendicular to the
line of sight, $\mathcal{Z}$-axis), we obtain a useful formula
\begin{equation}
\frac{G M_E}{R_E} = \frac{c^2}{4}
\frac{D_s}{D_{ls}} \theta_E~~~, \label{eq:einrad}
\end{equation}
which indicates that only mass inside the Einstein radius has
net effect on the deflection of light, regardless of the full lens mass distribution.

On the other hand, the measurement of
central velocity dispersion $\sigma$ can provide a
model-dependent dynamical estimate of this mass
based on the assumption of the power-law mass density profile $\rho$,
and luminosity density of stars $\nu$ \citep{Koopmans06}:
\begin{eqnarray}
\label{eq:rhopl}
\rho(r) &=& \rho_0 \left(\frac{r}{r_0}\right)^{-\alpha} \\
\nu(r) &=& \nu_0 \left(\frac{r}{r_0}\right)^{-\delta}
\label{eq:nupl}
\end{eqnarray}
Here $r$ is the spherical radial coordinate from the lens center and
it is related to the projected radial coordinate $R$ and the coordinate along the line of sight $\mathcal{Z}$ by:
 $r^2 = R^2 + \mathcal{Z}^2$. In order to
characterize the anisotropic distribution of three-dimensional
velocity dispersion, an anisotropy parameter $\beta$
is also introduced:
\begin{equation}
\beta(r) = 1 - {\sigma^2_t} / {\sigma^2_r} \label{eq:beta}
\end{equation}
where $\sigma_t$ and $\sigma_r$ are the tangential and radial
velocity dispersions, respectively. It has almost always been
assumed to be independent of $r$ \citep{Bolton06,Koopmans06}. In our
analysis we will consider two cases: isotropic distribution
$\beta=0$ and anisotropic distribution $\beta = const. \neq 0$. We
emphasize here that the power-law model adopted by us for the
luminosity profile is simply a convenient and flexible parameterized
mathematical model for early-type galaxies. In the previous work by
\citet{Bolton12}, it was verified that the results of mass-density
profile evolution do not depend significantly on the choice of the
specific form of parameterized luminosity-profile. Taking the Nuker
profile \citep{Lauer95}, a broken power-law profile including a
transition of variable softness between inner and outer regions, as
the reference model they found that differences between constraints
obtained with the Nuker and de Vaucouleurs \citep{Vaucouleurs48}
profiles were insignificant. However, in order to better describe
the distribution of luminous tracers in galaxies, more
data-orientated luminosity profiles like de Vaucouleurs, Hernquist,
Jaffe or Nuker profiles (see e.g.
\citet{Koopmans06a,Bolton12,Sonnenfeld13}) should be taken into
consideration in our next-step works.

Following the well-known spherical Jeans equation \citep{Binney80},
radial velocity dispersion of luminous matter $\sigma^2(r)$
of the early-type galaxies can be expressed as
\begin{equation}
\sigma^2_r(r) =  \frac{G\int_r^\infty dr' \ \nu(r') M(r') (r')^{2
\beta - 2} }{r^{2\beta} \nu(r)}~~~, \label{eq:binney}
\end{equation}
where $\beta$ is the above mentioned anisotropy parameter. Using
the mass density profile from Eq.~(\ref{eq:rhopl}), one can obtain the
relation between $M_E$ and the mass enclosed within a spherical radius $r$:
\begin{equation}
M(r) = \frac{2}{\sqrt{\pi} \lambda(\alpha)}
\left(\frac{r}{R_E}\right)^{3 - \alpha} M_E ~~~,
\end{equation}
where $\lambda(x) =
\Gamma \left(\tfrac{x-1}{2}\right) / \Gamma
\left(\tfrac{x}{2}\right)$ denotes the ratio of Euler's gamma functions.
Using the notation $\xi = \delta + \alpha - 2$, after
\citet{Koopmans06}, we obtain a convenient form for the radial
velocity dispersion by scaling the dynamical mass to the Einstein
radius:
\begin{equation}
\sigma^2_r(r) = \left[\frac{G M_E}{R_E} \right]
\frac{2}{\sqrt{\pi}\left(\xi- 2 \beta \right) \lambda(\alpha)}
\left(\frac{r}{R_E}\right)^{2 - \alpha}
\end{equation}

A key ingredient in all measurements concerning strong lensing systems is the {\em
observed} velocity dispersion, which is a projected, luminosity
weighted average of the radially-dependent velocity dispersion
profile of the lensing galaxy. In order to predict this value based
on a set of galaxy parameters, we start with Eq.~(\ref{eq:binney}). Note that this equation is valid when
the relationship between stellar number
density and stellar luminosity density is spatially constant, an
assumption unlikely to be violated appreciably within the effective
radius of the early-type lens galaxies under consideration.

Furthermore, the actual observed velocity dispersion is
measured over the spectrometer aperture $\theta_{\rm ap}$ blurred with
atmospheric seeing
In our analysis, we will apply the aperture weighting
function provided by \citet{Schwab10}
\begin{equation}
w(R) \approx  e^{-R^2/2 \tilde{\sigma}_{\rm atm}^2} ~~~,
\label{eq:wofr}
\end{equation}
where
\begin{equation}
\tilde{\sigma}_{\rm atm} \approx \sigma_{\rm atm} \sqrt{1 + \chi^2 /
4 + \chi^4 / 40}
\end{equation}
$\chi = \theta_{\rm ap} / \sigma_{\rm atm}$ and $\sigma_{\rm atm}$ is the seeing recorded
by the spectroscopic guide cameras during survey observations.
Considering the effects of aperture with atmospheric blurring and
luminosity-weighted averaging (see \citet{Schwab10} for details),
the observed velocity dispersion can be expressed as
\begin{eqnarray}
\nonumber \bar {\sigma}_*^2 &=& \left[\frac{c^2}{4}
\frac{D_s}{D_{ls}} \theta_E \right] \frac{2}{\sqrt{\pi}}
\frac{(2 \tilde{\sigma}_{\rm atm}^2/\theta_E^2)^{1-\alpha/2}}{ (\xi - 2\beta)} \\
&&\times\left[\frac{\lambda(\xi) - \beta \lambda(\xi+2)}
{\lambda(\alpha)\lambda(\delta)}\right] \frac{
\Gamma(\tfrac{3-\xi}{2}) }{\Gamma(\tfrac{3 - \delta}{2}) } ~~~.
\label{eq:plsig}
\end{eqnarray}

The observed stellar velocity dispersion is a luminosity-weighted
average dispersion inside the fiber aperture. Spectroscopic data
obtained in different surveys comprise luminosity averaged
line-of-sight velocity dispersions $\sigma_{ap}$ measured inside
different apertures. According to \citet{Cao15}, whose
compilation of lenses is used in this paper, SLACS and BELLS source
papers reported effective circular apertures, while for the SL2S and
LSD surveys they have been assessed from the sizes of the slit
reported in source papers. In our analysis, we take for
$\sigma_{atm}$ the median value recorded by the spectroscopic guide
cameras during survey observations. More specifically, we have added
$1''.4$ seeing for the SLACS spectroscopic observations according to
\citet{Bolton08} and $1''.8$ seeing for the BOSS spectroscopic
observations according to \citet{Bolton12, Shu15}. For the SL2S sample the seeing
for each individual lens was taken after \citet{Sonnenfeld13}. In the case of
LSD systems: CFRS03-1077, HST14176 and HST15433 the seeing data were taken from
\citet{Treu04}. Since they reported two seeing values per lens, we have taken the
median. For Q0047-281 we took the seeing value after \citet{Koopmans02}, but for MG2016+112 we assumed $0''.8$ seeing
because there was no seeing reported in the source paper.

From the above equation, one can immediately observe that there are
degeneracies among $\alpha$, $\beta$ and $\delta$. More importantly,
if we accurately knew the distance ratio $D_{s}/D_{ls}$, we could
get more stringent constraints on the parameters $\alpha$ and
$\delta$ describing the mass distribution of lensing galaxies. In
our analysis, the angular diameter distance $D_{12}(z)$ between
redshifts $z_1$ and $z_2$ (expressed in Mpc and assuming flat FRW
metric) is calculated as
\begin{eqnarray}
\label{inted} D_{12}=\frac{c}{H_0 (1+z_2)}\int_{z_1}^{z_2}
\frac{dz'}{E(z';\Omega_m)}
\end{eqnarray}
where $E(z; \Omega_m)= \sqrt{\Omega_m(1+z)^3+(1-\Omega_m)}$ is the
dimensionless Hubble function in flat $\Lambda$CDM model. We use the
best-fitted
matter density parameter $\Omega_m$ 
given by Planck Collaboration:
$\Omega_m=0.315$ 
\citep{Planck1}. Because our expressions involve only distance
ratios $D_{ls}/D_s$, the Hubble constant cancels and we do not have
to use its value, which according to Planck ($H_0 = 67.3 \;\; km \;
s^{-1} Mpc^{-1}$) was somewhat discrepant with alternative
estimates.

In order to determine $(\alpha, \delta)$ parameters of lensing
galaxies, we used Markov chain Monte Carlo (MCMC) method to
sample their probability density distributions based on the
likelihood ${\cal L} \sim \exp{(- \chi^2 / 2)}$, where
\begin{equation}
\chi^2 = \sum_{i=1}^{118} \left( \frac{ \bar
{\sigma}_{*,i}(z_{l,i},z_{s,i},\theta_{E,i}, \theta_{ap,i},
\sigma_{atm}; \alpha, \beta, \delta) - {\sigma}_{ap,i}}{\Delta \bar
{\sigma}_{*,i}} \right)^2
\end{equation}
was calculated using the measured values of
velocity dispersion $\sigma_{ap}$, Einstein radius $\theta_E$, and
aperture radius $\theta_{ap}$. Following the SLACS
team we took the fractional uncertainty of the Einstein radius at
the level of $5\%$, redshift measurements were assumed to be accurate.
The uncertainties of
$\sigma_{ap}$ and $\theta_E$ were propagated to the final uncertainty $\Delta \bar {\sigma}_{*}$ of
$\bar {\sigma}_{*}$. However, the
statistical error on Einstein radius was relatively small in
comparison to the velocity dispersion error.

In this paper, we used a comprehensive compilation of 118 strong
lensing systems observed by four surveys: SLACS, BELLS, LSD and
SL2S, which is also the largest gravitational lens sample (suitable
for the purpose of this study) published in our recent work
\citep{Cao15}. The SLACS data comprise 57 strong lenses presented
in \citet{Bolton08,Auger09}, the BELLS data comprise 25
lenses taken from \citet{Brownstein12}, then 5 most reliable lenses
from the LSD survey were taken after
\citet{Koopmans02,Treu02,Treu04}, and the SL2S data for a total of
31 lenses were taken from \citet{Sonnenfeld13a,Sonnenfeld13}.
Scatter plot showing the distribution of lenses from
different surveys in the redshift-velocity dispersion space can be
seen in Fig.~1 of \citet{Cao15}, from which one can see a fair
coverage of redshifts in the combined sample. We remind the readers
to refer to Table 1 of \citet{Cao15} for the full information about
all the 118 lenses. Because our list includes lensing galaxies
corresponding to different velocity dispersions at different
redshifts, so besides the full combined sample we also considered
six sub-samples separately. Within a singular isothermal
sphere model, the dynamical mass is related to the velocity
dispersion through the relation $M\propto\sigma^2$
\citep{Longair98}. Therefore, we considered three sub-samples defined by the velocity
dispersions of lenses \footnote{As a rough criterion, elliptical
galaxies with velocity dispersion smaller than 200 km/s are
classified as relatively low-mass galaxies, while those with
velocity dispersion larger than 300 km/s are treated as relatively
high-mass galaxies.}: $\sigma_{ap} \leq 200$ km/s ($n=25$ lenses),
$200$ km/s $< \sigma_{ap} \leq 300$ km/s ($n=80$ lenses), and
$\sigma_{ap} > 300$ km/s ($n=13$ lenses). Another set of three
sub-samples was defined by restriction to three redshift ranges
\footnote{In the full sample of 118 strong lenses, the highest lens
redshift of $z_l=1.00$ was recorded for the system MG2016. Therefore,
we set the breakpoints at 1/2 and 1/5 of the highest redshift, that is
at $z_l=0.50$ and $z_l=0.20$ respectively. We remark here that such an approach
does not represent any physical aspects of galaxy distribution in redshift, but
it guarantees that there are enough data points in
each sub-sample.}: $z\leq 0.20$ ($n=25$ lenses), $0.20<z\leq 0.50$
($n=65$ lenses), and $z>0.5$ ($n=28$ lenses).

\begin{table*}
\caption{\label{tab:result} Summary of constraints on the
galaxy structure parameters obtained with the full sample and six
sub-samples of strong lensing systems (see text for definitions).}
\begin{center}
\begin{tabular}{|l||llllllll}\hline\hline
Sample (Model)        & Galaxy structure parameters \{$\gamma$, $\alpha$, $\delta$\}  \\
\hline
Full sample ($\alpha=\delta=\gamma$)  & $\gamma_0= 2.132\pm 0.055$ (1$\sigma$), \ $\gamma_1= -0.067\pm 0.119$ (1$\sigma$)   \\
Full sample ($\alpha \neq \delta$)   & $\alpha_0= 2.070\pm0.031$ (1$\sigma$), \ $\alpha_1= -0.121\pm0.078$ (1$\sigma$), \\
                                     &  $\delta= 2.710\pm0.143$ (1$\sigma$)   \\
Full sample ($\alpha \neq \delta$)   & $\alpha= 2.035\pm0.013$ (1$\sigma$), \ $\delta= 2.681\pm0.164$ (1$\sigma$)   \\
\hline

Sub-sample ($z\leq 0.2$) ($\alpha=\delta=\gamma$) & $\gamma_0= 2.175^{+0.105}_{-0.218}$ (1$\sigma$), \ $\gamma_1= -0.495^{+1.345}_{-0.565}$ (1$\sigma$)\\
Sub-sample ($0.2<z\leq 0.5$) ($\alpha=\delta=\gamma$) & $\gamma_0= 2.093\pm 0.114$ (1$\sigma$), \ $\gamma_1= 0.063\pm 0.307$ (1$\sigma$)  \\
Sub-sample ($z>0.5$) ($\alpha=\delta=\gamma$)   & $\gamma_0= 2.275\pm 0.269$ (1$\sigma$), \ $\gamma_1= -0.288\pm 0.394$ (1$\sigma$)    \\
\hline

Sub-sample ($\sigma_{ap} \leq 200$ km/s) ($\alpha=\delta=\gamma$)  & $\gamma_0= 2.135\pm 0.087$ (1$\sigma$), \ $\gamma_1= 0.012\pm 0.204$ (1$\sigma$)  \\
Sub-sample ($200$ km/s $< \sigma_{ap} \leq 300$ km/s) ($\alpha=\delta=\gamma$) & $\gamma_0= 2.115\pm 0.072$ (1$\sigma$), \ $\gamma_1= -0.091\pm 0.154$ (1$\sigma$)   \\
Sub-sample ($ \sigma_{ap} > 300$ km/s) ($\alpha=\delta=\gamma$) & $\gamma_0= 1.982\pm 0.154$ (1$\sigma$), \ $\gamma_1= -0.047\pm 0.350$ (1$\sigma$)  \\
\hline

Sub-sample ($z\leq 0.2$) ($\alpha \neq \delta$) & $\alpha= 2.067^{+0.077}_{-0.140}$ (1$\sigma$), \ $\delta= 2.410^{+0.410}_{-0.242}$ (1$\sigma$)\\
Sub-sample ($0.2<z\leq 0.5$) ($\alpha \neq \delta$) &$\alpha= 2.029\pm0.018$ (1$\sigma$), \ $\delta= 2.700\pm0.180$ (1$\sigma$)   \\
Sub-sample ($z>0.5$) ($\alpha \neq \delta$)   & $\alpha= 1.982\pm0.044$ (1$\sigma$), \ $\delta= 2.731\pm0.216$ (1$\sigma$)   \\
\hline

Sub-sample ($\sigma_{ap} \leq 200$ km/s) ($\alpha \neq \delta$)  & $\alpha= 1.754\pm0.179$ (1$\sigma$), \ $\delta= 2.660\pm 0.120$ (1$\sigma$)  \\
Sub-sample ($200$ km/s $< \sigma_{ap} \leq 300$ km/s) ($\alpha \neq \delta$) & $\alpha= 2.004\pm0.065$ (1$\sigma$), \ $\delta=2.539\pm0.279$ (1$\sigma$)   \\
Sub-sample ($ \sigma_{ap} > 300$ km/s) ($\alpha \neq \delta$) & $\alpha= 1.829\pm 0.190$ (1$\sigma$), \ $\delta= 2.101\pm 0.149$ (1$\sigma$)  \\

\hline\hline
\end{tabular}
\end{center}
\end{table*}

\section{Results and discussions}\label{sec:results}

In this paper, we focused on constraining the parameters ($\alpha$,
$\delta$) characterizing the structure of elliptical galaxies using
different samples of strong lensing systems, i.e. the full $n=118$
sample as well as six sub-samples defined with different selection
criteria. These parameters are assumed to be the same for all
lenses, and we used MCMC method based on the publicly
available CosmoMC package \citep{Lewis02} to sample their
probability density distributions. To be specific, we generated
eight chains and stopped sampling when the worst e-value [the
variance(mean)/ mean(variance) of 1 / 2 chains] $R-1$ is of the
order 0.01.

\subsection{The case with $\alpha=\delta=\gamma$}

\begin{figure}
\begin{center}
\includegraphics[width=1.0\hsize]{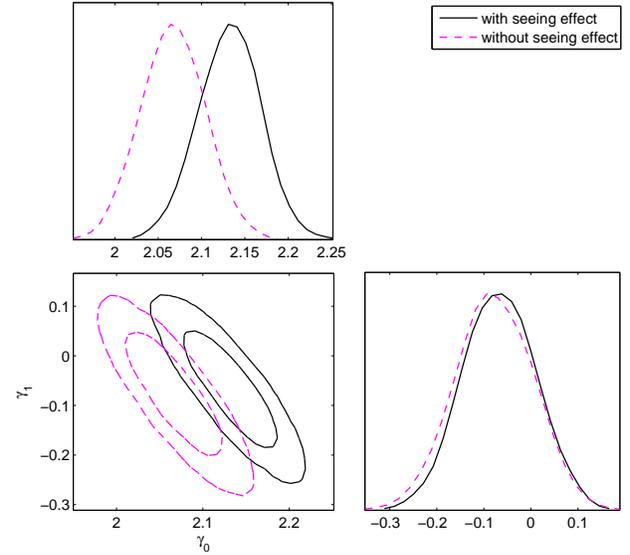}
\end{center}
\caption{Constraints on the total-mass density parameter obtained
from the full sample of strong lensing systems (Solid line). Limits
in the plane of $\gamma_0$ and $\gamma_1$ without using the seeing
observations are also shown for comparison (Dashed
line).\label{fig1}}
\end{figure}

In the first case analysed, we assume that the radial profile of the
luminosity density ($\nu$) follows that of the total mass density
($\rho$), i.e., $\alpha=\delta=\gamma$, and both of them evolve as a
function of redshift:
\begin{equation}
\gamma(z) =\gamma_{0}+\gamma_1z,
\end{equation}
where $\gamma_{0}$ is the present-day value and $\gamma_{1}$
characterizes the evolution of $\gamma$ with redshit. In order to
make comparison with the previous results, we assume that stellar
velocity anisotropy vanishes $\beta=0$. Using the full data-set, we
obtain the following best-fit values and corresponding 1$\sigma$
uncertainties
\begin{eqnarray}
&& \gamma_0= 2.132\pm 0.055, \nonumber\\
&& \gamma_1= -0.067\pm 0.119. \nonumber
\end{eqnarray}
Marginalized 1$\sigma$ and 2$\sigma$ contours for each parameter are
shown in Fig.~\ref{fig1}. The results reveal compatibility between
our full lens sample and a smaller combined sample from SLACS, SL2S
and LSD used previously by \citet{Ruff11,Sonnenfeld13}. Our results
suggest that the total density profile of early-type galaxies has
become slightly steeper over cosmic time ($z\sim 1$), which might
indicate the effect of dissipative processes in the growth of
massive galaxies. It is interesting to see the difference in the final results
obtained with and without seeing taken into account.
Therefore, Fig.~\ref{fig1} also shows the the limits in the plane of $\gamma_0$
and $\gamma_1$ based on the full sample of lenses without accounting for seeing
(see \citet{Koopmans06} for details).
In such a case, we have $\gamma_0= 2.066\pm
0.057$ and $\gamma_0= -0.077\pm 0.126$, which is consistent with the singular isothermal sphere ($\gamma_0=2$,
$\gamma_1=0$) model within $1\sigma$.

\begin{figure}
\begin{center}
\includegraphics[width=1.0\hsize]{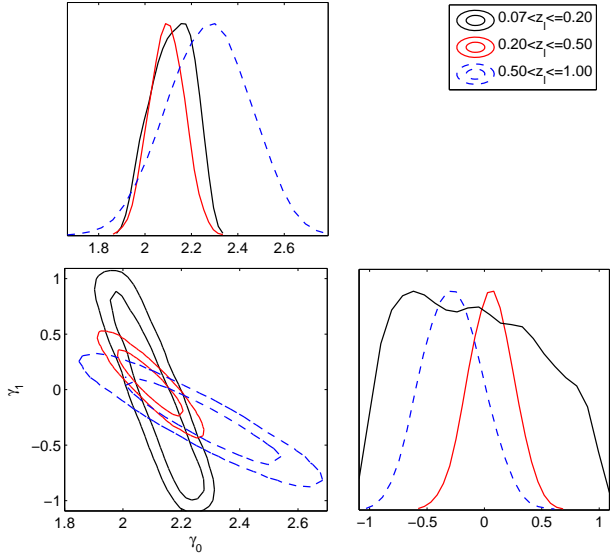}
\end{center}
\caption{Constraints on the total-mass density parameter obtained
from sub-sample of strong lenses defined by three different redshift bins.
\label{fig2}}
\end{figure}

\begin{figure}
\begin{center}
\includegraphics[width=1.0\hsize]{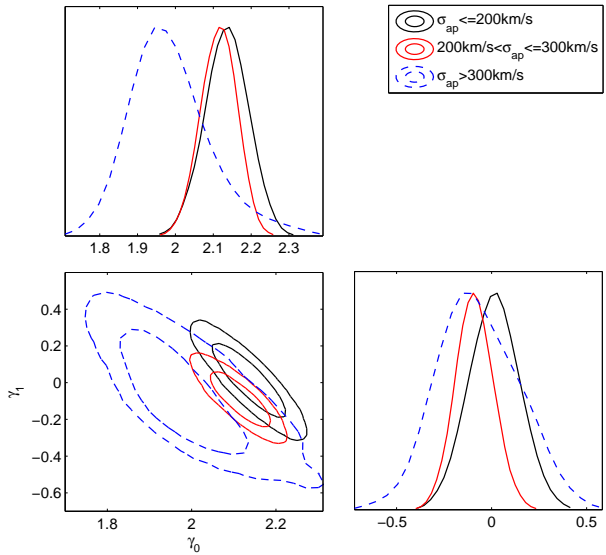}
\end{center}
\caption{Constraints on the total-mass density parameter obtained
from three sub-samples of strong lenses selected according to their velocity dispersions.
\label{fig3}}
\end{figure}

In Table 1 and Fig.~\ref{fig2}-\ref{fig3}, we also show the results
for $\gamma_0$ and $\gamma_1$ obtained on six sub-samples described
in Section~\ref{sec:data}. Concerning the sub-samples with different
lens redshift bins, we find that constraints coming from the three
sub-samples are in perfect agreement with each other. It is good to
recall \citep{Cao15} the median values of the lens redshifts for
different surveys: SLACS -- $z_l = 0.215$, BELLS -- $z_l = 0.517$,
LSD -- $z_l = 0.81$ and SL2S -- $ z_l = 0.456$. Note that SL2S
survey would be particularly promising in the future since it has
already reached the maximum lens redshift of $z_l = 0.80$. For other
three sub-samples of lenses differing by velocity dispersions (which
is directly related to the masses of galaxies, $M\propto \sigma^2$),
we note that the ranges of $\gamma$ parameters for relatively
low-mass galaxies ($\gamma_0=2.135\pm 0.087$, $\gamma_1=0.012\pm
0.204$) are close to estimates obtained for intermediate-mass
elliptical galaxies ($\gamma_0=2.115\pm 0.072$, $\gamma_1=-0.091\pm
0.154$). The best-fit values of $\gamma_0$ and $\gamma_1$ for
massive galaxies ($\gamma_0= 1.982\pm 0.154$, $\gamma_1= -0.047\pm
0.350$) are significantly different from the corresponding
quantities of low-mass and intermediate-mass elliptical galaxies. On
the other hand, the singular isothermal sphere model (SIS) is only
consistent with results obtained in massive (high velocity
dispersion) galaxies. The SIS value of $\gamma=2$ is indeed very
close to the central fit values. Consequently, our results imply
that the mass-density profile is different in low, intermediate and
high-mass elliptical galaxies. This implies the need of treating
these classes of galaxies separately.

\subsection{The case with $\alpha \neq \delta$}

\begin{figure}
\begin{center}
\includegraphics[width=1.0\hsize]{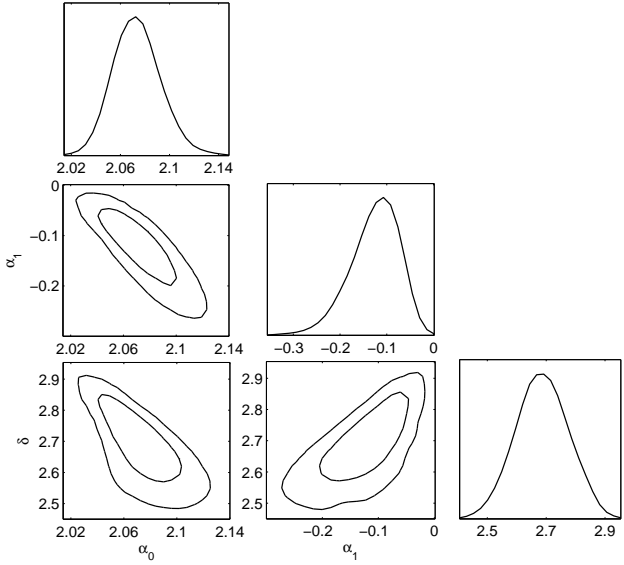}
\end{center}
\caption{ Constraints on the total-mass and luminosity
density parameters obtained from the full sample of strong lensing
systems. Redshift evolution of the mass-density slope is
parameterized as $\alpha=\alpha_0+\alpha_1z$. \label{fig4} }
\end{figure}

In the second case, we allow the luminosity density profile to be
different from the total-mass density profile, i.e., $\alpha \neq
\delta$, and the stellar velocity anisotropy exits, i.e., $\beta
\neq 0$. Moreover, we characterize anisotropy $\beta$ by a Gaussian
distribution, $\beta=0.18\pm0.13$, based on the well-studied sample
of nearby elliptical galaxies from \citet{Gerhard01}. The above
cited uncertainty $\sigma_{\beta} = 0.13$ represents the intrinsic spread of
this quantity \citep{Bolton06,Schwab10}. In order to
alleviate the degeneracy between luminosity density and total-mass
density profiles, which might yield more meaningful results, we
firstly include the redshift evolution of the mass-density slope
$\alpha=\alpha_0+\alpha_1z$ where $\alpha$ is not equal to $\delta$
(the luminosity-density slope). Performing fits on the full
data-set, the 68\% confidence level uncertainties on the three model
parameters are
\begin{eqnarray}
&& \alpha_0= 2.070\pm0.031, \nonumber\\
&& \alpha_1= -0.121\pm0.078, \nonumber\\
&& \delta= 2.710\pm0.143. \nonumber
\end{eqnarray}
Fig.~\ref{fig4} shows these constraints in the parameter space of
$\alpha_0$, $\alpha_1$ and $\delta$. It is interesting to note that
the values of mass-density exponents are inconsistent with the
values obtained in the case of $\alpha = \delta$. This implies that
from the point of view of stellar dynamics, the effective
description of mass distribution in lensing galaxies can be much
different when $\alpha \neq \delta$. Moreover, the obtained value of
$\delta$ is consistent with that of \citet{Schwab10}, which fit the
PSF convolved two-dimensional power-law ellipsoid images of 53 SLACS
lenses to their corresponding HST F814W imaging data within a circle
of radius \citep{Bolton06}. However, we also notice that all of these
three parameters have relatively large uncertainties. In order to
obtain more stringent results, we will further ignore the redshift
evolution of the mass-density slope $\alpha_1=0$.

\begin{figure}
\begin{center}
\includegraphics[width=1.0\hsize]{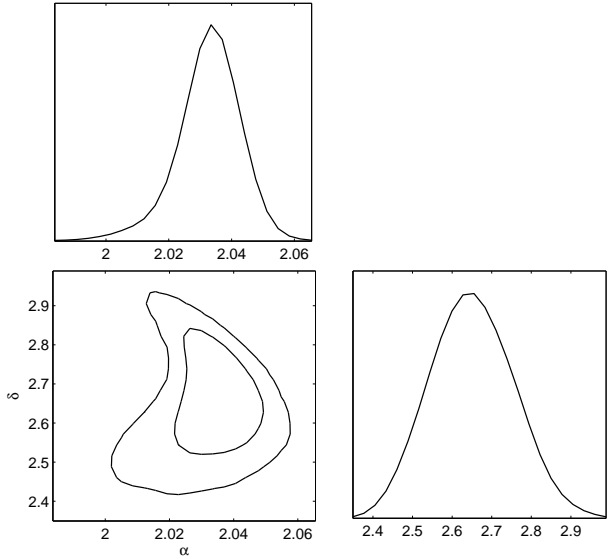}
\end{center}
\caption{Constraints on the total-mass and luminosity density
parameters obtained from the full sample of strong lensing systems.
\label{fig5}}
\end{figure}

Performing fits on the full data-set, we obtain the following
best-fit values and corresponding 68\% confidence level
uncertainties
\begin{eqnarray}
&& \alpha= 2.035\pm0.013, \nonumber\\
&& \delta= 2.681\pm0.164. \nonumber
\end{eqnarray}
From the results displayed in Fig.~\ref{fig5}, one can see that the
power-law exponent $\delta$, which describes the luminosity density
profile of elliptical galaxies, takes values from the range which is
different from the range of mass-density exponent $\alpha$. Our
result is also in good agreement with a recent analysis of lensing
statistics by \citet{Schwab10}, which reported mean value of
$\avg{\delta} = 2.40$ and a standard deviation $\sigma_\delta =
0.11$. The different range of $(\alpha, \delta)$ parameters reveals
differences in mass density distributions of dark matter and
luminous baryons in early-type galaxies. A model in which mass
traces light ($\alpha=\delta$) is rejected at $>95\%$ confidence and
our analysis robustly indicates the presence of dark matter in the
form of a mass component that is differently spatially distributed
than stars.

\begin{figure}
\begin{center}
\includegraphics[width=1.0\hsize]{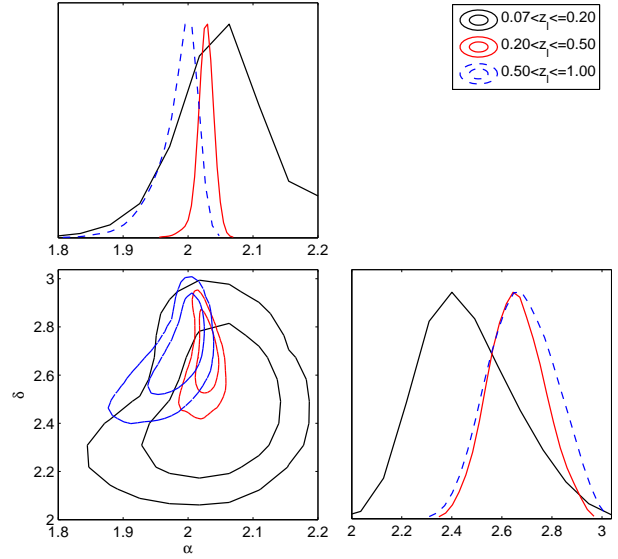}
\end{center}
\caption{Constraints on the total-mass and luminosity density
parameters obtained from sub-samples of strong lensing systems
defined by three different redshift bins. \label{fig6}}
\end{figure}

\begin{figure}
\begin{center}
\includegraphics[width=1.0\hsize]{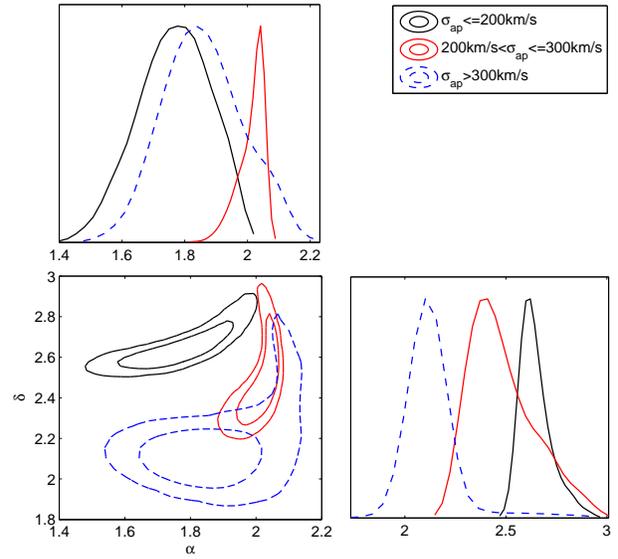}
\end{center}
\caption{Constraints on the total-mass and luminosity density
parameters obtained from three sub-samples of strong lensing systems
with different velocity dispersions. \label{fig7}}
\end{figure}

From the results obtained on six sub-samples shown in
Fig.~\ref{fig6}-\ref{fig7}, one can see again the consistency
between three sub-samples defined according to the lens redshift
bins. Concerning the sub-samples differing by velocity dispersion
(i.e. by mass as well) both the total-mass and luminosity
distributions exhibit different density profiles across sub-samples.
The constraint results on $\alpha$ parameter are particularly
interesting. Namely, one can see that samples of elliptical galaxies
differing by mass have different power-law profiles of total-mass
density distribution: $\alpha= 1.754\pm0.179$ for $\sigma_{ap} \leq
200$ km/s, $\alpha= 2.004\pm0.065$ for $200$ km/s $< \sigma_{ap}
\leq 300$ km/s, and $\alpha= 1.829\pm 0.190$ for $\sigma_{ap} > 300$
km/s. Substantial distinction between $\alpha$ and $\delta$
parameters exists for all three sub-populations within $1\sigma$. It
is of interest to note that the intermediate-mass elliptical
galaxies are consistent with the singular isothermal sphere within
$1\sigma$ region. Taking the luminosity profile of elliptical
galaxies into consideration, the obtained value of $\delta$ from our
sub-sample with intermediate velocity dispersion ($200$ km/s $<
\sigma_{ap} \leq 300$ km/s), whose confidence contours in $(\alpha,
\delta)$ parameter plane differ the other two remaining samples, is
in perfect agreement with the previous results from smaller SLACS
sample \citep{Schwab10}.

Previously, some researchers \citep{Auger10, Sonnenfeld13,
Dutton2014} have considered the problem of how does the mass
weighted slope within the effective radius correlate with other
properties of strong lenses like their stellar density or the
velocity dispersion. In particular \citet{Dutton2014} compared the
observed trends with calculated expectations based on different
evolutionary scenarios for early type galaxies, such like adiabatic
halo contraction, mild or no contraction or halo expansion. The
trends of slope factors with the velocity dispersion found in our
study (Table~1), in the case of $\alpha=\delta$ are in rough
agreement with the adiabatic contraction scenario. We stress,
however that the uncertainties in estimated parameters do not allow
us to discriminate between alternative behaviors. In the case of
$\alpha \neq \delta$ we found no trend in central fit values --
slope steepens between low-mass and intermediate-mass galaxies and
then again becomes more shallow for massive galaxies. This
observation should be taken with caution. First because these fits
agree with each other within $1\sigma$ uncertainty. Second, because
this case is hard to compare quickly against the results obtained by
the others who adopted Sersic or de Vaucouleurs profiles for the
light.

\section{Conclusion}

In this paper, we explored 118 strong gravitational lenses observed
by SLACS, BELLS, LSD and SL2S surveys to constrain the total
mass-profile and light-profile shapes of elliptical galaxies since
redshift $z \sim 1$. Our method of statistical analysis is the same
as in Cao et al.(2015). However, we used as a prior the best-fitted
$\Lambda$CDM cosmology from Planck and assumed power-law density
profiles for the total mass density, $\rho=
\rho_0(r/r_0)^{-\alpha}$, and luminosity density, $\nu=
\nu_0(r/r_0)^{-\delta}$. Allowing for the evolution we also
investigated the total mass density profile exponent and its first
derivative with respect to the redshift in the form of
$\alpha=\alpha_0+\alpha_1z$. First, we assumed, as it was done
previously by the others, that light tracers and total mass follow
the same profile $\alpha=\delta=\gamma$. Then, using the full
sample, we obtained $\gamma_0= 2.132\pm 0.055$ and a mild trend
$\gamma_1= -0.067\pm 0.119$, suggesting that the total density
profile of massive galaxies has become slightly steeper over cosmic
time. Furthermore, we divided the full sample into six different
sub-samples according to the lens redshifts and their velocity
dispersions respectively. It turned out that there are no
significant differences between lenses from different redshift bins.
It is perhaps due to fact that the redshift range covered by lenses
is not big enough to display any noticeable differences. However,
the division according to velocity dispersion (i.e. effectively
according to mass) turned out to be more discriminative. Low and
intermediate mass galaxies show similar profiles ($\gamma_0=2.135\pm
0.087$, $\gamma_1=0.012\pm 0.204$ and $\gamma_0=2.115\pm 0.072$,
$\gamma_1=-0.091\pm 0.154$, respectively), while the best-fit values
of $\gamma_0$ and $\gamma_1$ for massive galaxies ($\gamma_0=
1.982\pm 0.154$, $\gamma_1= -0.047\pm 0.350$) are significantly
different from the corresponding quantities of other two
sub-samples. The singular isothermal sphere model (SIS) is only
consistent with results obtained in massive (high velocity
dispersion) galaxies. Consequently, our results imply that the total
mass density profiles of intermediate-mass and high-mass elliptical
galaxies are different.

Then, we set $\delta$ as a free parameter and allow the luminosity
density profile to be different from the total-mass density profile,
i.e., $\alpha \neq \delta$. Performing fits on the full data and
considering the redshift evolution of $\alpha=\alpha_0+\alpha_1z$,
we obtained $\alpha_0= 2.070\pm0.031$, $\alpha_1= -0.121\pm0.078$,
and $\delta= 2.710\pm0.143$. This value is also in good agreement
with a recent analysis by \citep{Schwab10} obtained on a much
smaller sample. A model in which mass traces light (i.e.
$\alpha=\delta$) is rejected at $>95\%$ confidence and our analysis
robustly indicates the presence of dark matter in the form of a mass
component distributed differently from the light. Fits performed on
six sub-samples lead to the similar conclusion as in previous case.
In particular, the substantial distinction between $\alpha$ and
$\delta$ admissible ranges exists in all sub-samples. It is
interesting to note that only intermediate-mass elliptical galaxies
are consistent with the singular isothermal sphere within $1\sigma$,
i.e, $\alpha= 2.004\pm0.065$. Taking the luminosity profile of
elliptical galaxies into consideration, the obtained value of
$\delta$ from our sub-sample with intermediate velocity dispersion
is in perfect agreement with the previous results from smaller SLACS
sample \citep{Schwab10}.

As a final remark, we point out that the sample discussed in this
paper is based on strong lensing systems discovered in different
surveys. Our analysis potentially may suffer from systematics
stemming from this inhomogeneity. If the best fitted values
of $\alpha$ and $\delta$ power-law exponents are robustly confirmed
in future larger samples, they can be used to elaborate a
more accurate phenomenological model of elliptical galaxies. Such a
model going beyond the SIS, would serve as a more realistic prior
assumption in cosmological tests based on distance ratios.

\section*{Acknowledgments}
We are grateful to the referee for very constructive discussion and
useful comments that allowed us to improve the paper considerably.
This work was supported by the Ministry of Science and Technology
National Basic Science Program (Project 973) under Grants Nos.
2012CB821804 and 2014CB845806, the Strategic Priority Research
Program ``The Emergence of Cosmological Structure" of the Chinese
Academy of Sciences (No. XDB09000000), the National Natural Science
Foundation of China under Grants Nos. 11503001, 11373014 and
11073005, the Fundamental Research Funds for the Central
Universities and Scientific Research Foundation of Beijing Normal
University, China Postdoctoral Science Foundation under grant No.
2015T80052, and the Opening Project of Key Laboratory of
Computational Astrophysics, National Astronomical Observatories,
Chinese Academy of Sciences. Part of the research was conducted
within the scope of the HECOLS International Associated Laboratory,
supported in part by the Polish NCN grant DEC-2013/08/M/ST9/00664 -
M.B. gratefully acknowledges this support. This research was also
partly supported by the Poland-China Scientific \& Technological
Cooperation Committee Project No. 35-4. M.B. obtained approval of
foreign talent introducing project in China and gained special fund
support of foreign knowledge introducing project.


\label{lastpage}

\end{document}